\documentclass[pre,aps,twocolumn,showpacs,superscriptaddress,nofootinbib]{revtex4}
\usepackage{graphicx} 
\usepackage{dcolumn}  
\usepackage{bm}       
\usepackage{amsmath}
\usepackage{epsfig}

\def\ii{{\rm i}}  \def\ee{{\rm e}}    
    \def\pb{{\bf p}}  
    \def\Eb{{\bf E}}  \def\rb{{\bf r}}
\def\xx{\hat{\bf x}}   
\def\rhoh{{\hat{\bf \rho}}}  \def\phih{{\hat{\bf \varphi}}}
\def\yy{\hat{\bf y}}     
\def\zz{\hat{\bf z}}

\begin{document}
\title{Stimulated light emission and inelastic scattering by a classical linear system of rotating particles}
\author{Ana Asenjo-Garcia}
\affiliation{Instituto de \'Optica - CSIC, Serrano 121, 28006
Madrid, Spain}
\author{Alejandro Manjavacas}
\affiliation{Instituto de \'Optica - CSIC, Serrano 121, 28006
Madrid, Spain}
\author{F. Javier Garc\'{\i}a de Abajo}
\email[Corresponding author: ]{J.G.deAbajo@csic.es}
\affiliation{Instituto de \'Optica - CSIC, Serrano 121, 28006
Madrid, Spain}
\affiliation{Optoelectronics Research Centre, University of Southampton, Southampton SO17 1BJ, UK}

\date{\today}

\begin{abstract}
The rotational dynamics of particles subject to external illumination is found to produce light amplification and inelastic scattering at high rotation velocities. Light emission at frequencies shifted with respect to the incident light by twice the rotation frequency dominates over elastic scattering within a wide range of light and rotation frequencies. Remarkably, net amplification of the incident light is produced in this classical linear system via stimulated emission. Large optically-induced acceleration rates are predicted in vacuum accompanied by moderate heating of the particle, thus supporting the possibility of observing these effects under extreme rotation conditions.
\end{abstract}
\pacs{42.50.Wk,41.60.-m,45.20.dc,78.70.-g}
\maketitle



\section{Introduction}

Despite its minuteness, the exchange of momentum by light-matter interaction is responsible for observable phenomena ranging from the formation of comet tails to the trapping \cite{A1970,G03} and cooling \cite{GHV09} of small particles down to single atoms. In particular, the exchange of angular momentum produces mechanical torques. In a pioneering experiment, Beth showed that optical spin can induce rotation in birefringent plates illuminated by circularly polarized (CP) light \cite{B1936}. This effect has been subsequently confirmed on the macro- \cite{A1966}, micro- \cite{FNH98}, and nanoscales \cite{LZL10,K10}. Light can also carry orbital angular momentum, which has been widely used to produce vortex-like motion of microparticles \cite{GS98}. Additionally, chiral particles undergo rotation even when illuminated by unpolarized plane waves \cite{GO01,TMK10} (i.e., light without net angular momentum).
 
The interaction of light with rotating particles raises fundamental questions, such as the possibility of cooling the rotational degrees of freedom down to the quantum regime. On the opposite side, the material response of particles rotating at extreme velocities can be largely influenced by spinning forces, eventually leading to centrifugal explosion. Even the kinematical change between lab and rotating frames is known to produce frequency shifts in the light emitted by rotating particles \cite{BB97,CDR98,CRD98,MHS05}. These effects may be occurring in cosmic dust irradiated by polarized light over enormous periods of time. Particle trapping in vacuum \cite{AD1976} may provide a suitable framework to study these phenomena.

In this Letter, we study the electromagnetic torque and scattering properties of rotating particles suject to external illumination. Like in rotational Raman scattering \cite{G1981}, the particle produces inelastic scattering at frequencies separated from the incoming light by twice the rotation frequency. We report two remarkable observations at large rotation velocities: (i) inelastic scattering is stronger than elastic scattering; and (ii) net amplification of the incident light takes place via stimulated emission in this purely classical linear system. We base this conclusions on analytical expressions derived for the torque, the absorbed power, and the inelastic light scattering cross section, from which a complex resonant interplay between the light frequency and the particle rotation frequency is observed. Feasible experimental conditions for the observation of light amplification are discussed.

\begin{figure}
\begin{center}
\includegraphics[width=60mm,angle=0,clip]{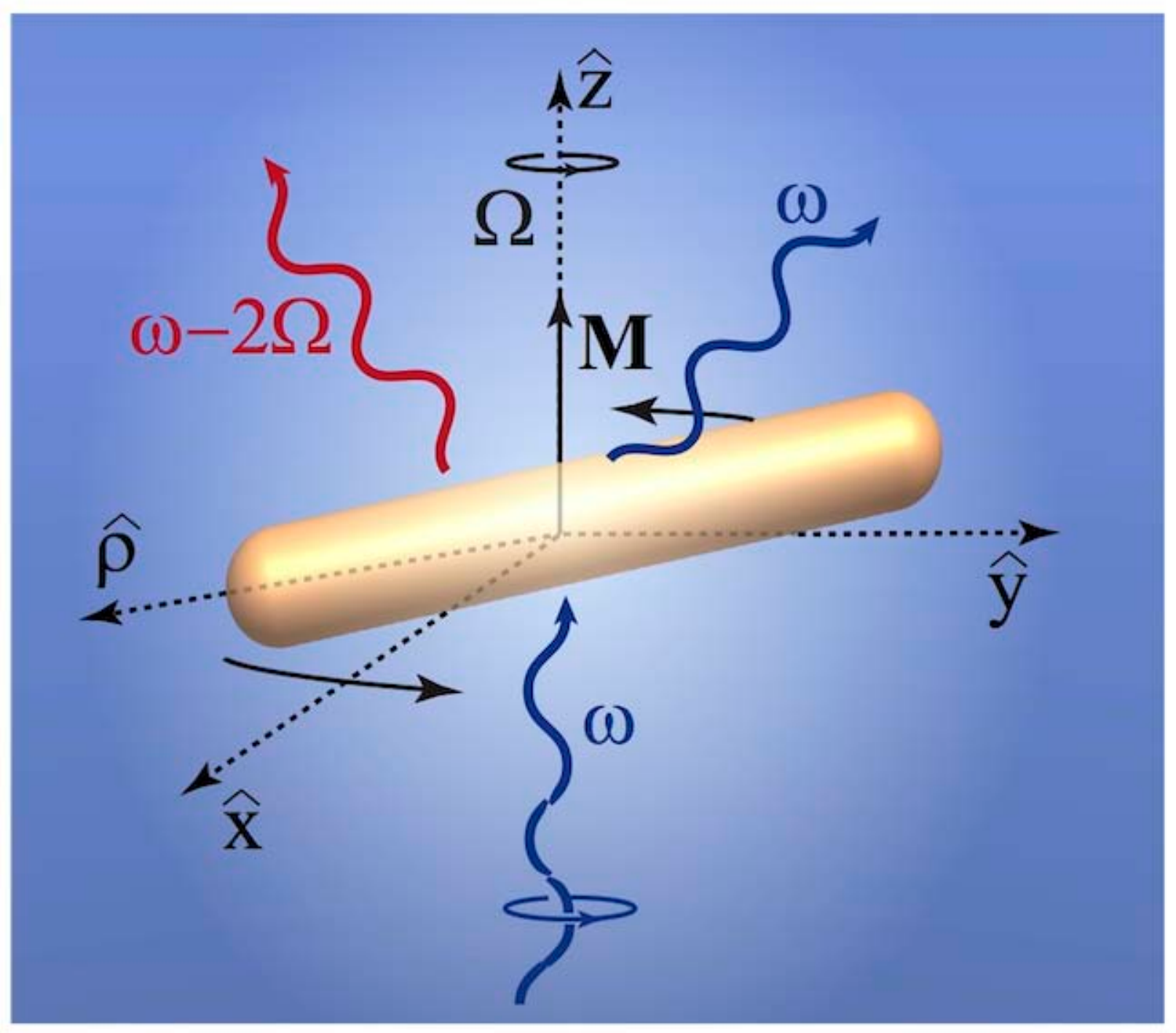}
\caption{Sketch of a rod-like particle rotating with frequency $\Omega$ around a transversal direction ($z$ axis). An incident light plane wave of frequency $\omega$ illuminates the rod along a direction normal to its axis, thus producing an electromagnetic torque $M$ and scattered light of frequencies $\omega$ and $\omega\pm2\Omega$. For left circularly polarized (LCP) light, as considered in the figure, only the $\omega-2\Omega$ inelastic component is emitted.} \label{Fig1}
\end{center}
\end{figure}

\section{Oscillator model for a spinning particle}

For simplicity, we consider a spinning rod, the optical response of which can be modeled with an effective spring consisting of a point particle of mass $m$ and charge $Q$ oscillating around a fixed charge $-Q$ with natural frequency $\omega_0$ and damping rate $\gamma$. The point-particle motion is constrained to the longitudinal direction of the rod $\rhoh$, which is rotating with frequency $\Omega$ around a direction $\zz$ perpendicular to $\rhoh$ (see Fig.\ \ref{Fig1}), so that the instantaneous position of this charge can be written $\rb(t)=\rho(t)\left(\xx\cos\Omega t+\yy\sin\Omega t\right)$, where $\rho(t)$ is the distance to the central charge. The force equation of motion is then given by (see Appendix)
\begin{eqnarray}
m\ddot{\rb}=-m\omega_0^2\rho\rhoh-m\gamma\dot{\rho}\rhoh+Q\Eb+m\tau\dddot{\rb}+F^{\rm react}\phih,
\label{force}
\end{eqnarray}
where the terms on the right-hand side are (from left to right) the restoring force, the intrinsic friction, the external electric-field force, the Abraham-Lorentz force accounting for radiative damping proportional to $\tau=2Q^2/3mc^3$ \cite{J99}, and the particle reaction force $F^{\rm react}$ taken to cancel other azimuthal components in order to constrain the motion of the point charge along the rotating $\rhoh$ axis. We assume the charge velocity to be small compared to the speed of light, so that the magnetic component of the force can be overlooked, although it could lead to a small precession. The transversal polarization of the rod is also considered to be negligible.

\begin{figure*}
\begin{center}
\includegraphics[width=170mm,angle=0,clip]{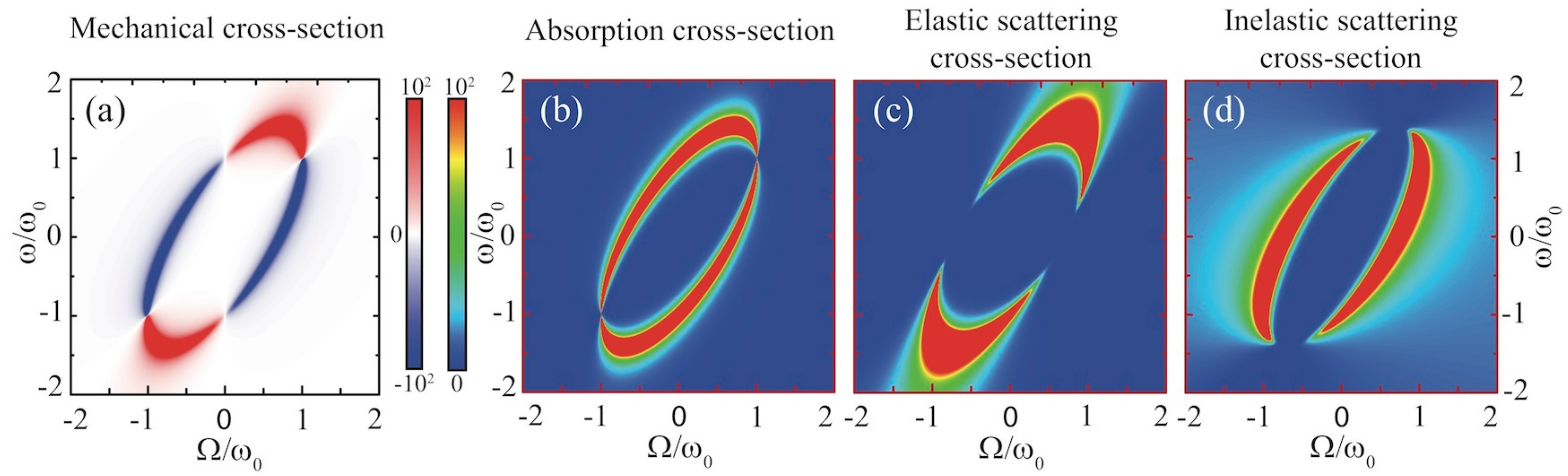}
\caption{Dynamical properties of a rod illuminated by LCP light of frequency $\omega$ and rotating at frequency $\Omega$, both normalized to the excitation resonance frequency of the rod $\omega_0$. {\bf (a)} Mechanical cross section $M\Omega/I$, where $M$ is the electromagnetic torque and $I$ is the light intensity. {\bf (b)} Absorption cross-section. {\bf (c)} Elastic scattering cross-section. {\bf (d)} Inelastic scattering cross-section (emission at frequency $\omega-2\Omega$). The particle is described by a rotating harmonic oscillator with intrinsic damping rate $\gamma=0.1\omega_0$. The color scale is in units of $Q^2\gamma/mc\omega_0^2$ in (a)-(b) and $Q^2\tau/mc$ in (c)-(d), assuming $\gamma\gg\tau\omega_0^2$.} \label{Fig2}
\end{center}
\end{figure*}

\section{Electromagnetic torque}

The resulting torque $M$ is directed along $\zz$ and originates in the reaction force according to $M=-F^{\rm react}\rho$. For monochromatic incident light of frequency $\omega$ and electric field $\Eb=(E_x\xx+E_y\yy)\ee^{-\ii\omega t}+c.c.$, we can solve Eq.\ (\ref{force}) analytically to find the time-averaged torque (see Appendix)
\begin{align}
M=&\frac{Q^2}{2m}\Bigg\{-\left[\gamma\omega_++\tau(\omega+2\Omega)^3\right]
\frac{|E_+|^2}{|d_+|^2}\nonumber\\ &+\left[\gamma\omega_-+\tau(\omega-2\Omega)^3\right]
\frac{|E_-|^2}{|d_-|^2}\Bigg\},
\label{M}
\end{align}
where $\omega_\pm=\omega\pm\Omega$, $E_\pm=E_x\pm\ii E_y$, and $d_\pm=\omega_0^2-\Omega^2-\omega_\pm(\omega_\pm+\ii\gamma)-\ii\tau\omega_\pm(\omega_\pm^2+3\Omega^2)$.
The $\gamma$ terms in Eq.\ (\ref{M}) originate in photon absorption by the particle, whereas the $\tau$ terms describe a radiative reaction torque similar to what happens in rotating dipoles \cite{C1967}. The torque predicted by Eq.\ (\ref{M}) presents resonant features signaled by the zeros of $d_\pm$. In small absorbing particles, for which the $\tau$ terms can be neglected compared to the $\gamma$ absorption terms, this condition reduces to the ellipses
\begin{eqnarray}
\Omega^2+(\omega\pm\Omega)^2=\omega_0^2,
\label{cond}
\end{eqnarray}
as is clear in Fig.\ \ref{Fig2}(a) for left CP (LCP) incident light, corresponding to the lower sign in Eq.\ (\ref{cond}). The figure is actually representing the cross section towards mechanical work, given by $\sigma_{\rm mech}=M\Omega/I$, where $I=(c/2\pi)|E|^2$ is the external light intensity and $|E|^2=|E_x|^2+|E_y|^2$. For light oscillating at the same frequency as the rotation, the particle sees a frozen incident field, and the torque is therefore zero (it vanishes along the $\omega=\Omega$ line). For slower rotation ($\omega>\Omega$) the torque is positive, pointing to a net transfer of momentum from the light field to the particle. In contrast, the torque and $\sigma_{\rm mech}$ become negative for faster rotation even when the light and the particle rotate in the same direction ($\Omega>\omega>0$). This suggests that the incident light is actually braking the particle by effectively producing stimulated photon emission with contributions proportional to the ohmic and radiative losses (terms in $\gamma$ and $\tau$, respectively).

\section{Partial cross sections}

The absorption and elastic-scattering cross sections admit the closed-form expressions (see Appendix)
\begin{eqnarray}
\sigma_{\rm abs}=\frac{\pi\gamma Q^2}{mc}\left(\omega_+^2
\frac{|E_+/E|^2}{|d_+|^2}+\omega_-^2\frac{|E_-/E|^2}{|d_-|^2}\right)
\nonumber 
\end{eqnarray}
and
\begin{eqnarray}
\sigma_\omega=\frac{\pi Q^4\omega^4}{3m^2c^4}\left(\frac{|E_+/E|^2}{|d_+|^2}+\frac{|E_-/E|^2}{|d_-|^2}\right).
\nonumber 
\end{eqnarray}
Interestingly, the rotational motion produces inelastic scattering components at frequencies $\omega\pm2\Omega$ \cite{G1981} and responding to the cross sections (see Appendix)
\begin{eqnarray}
\sigma_{\omega\pm2\Omega}=\frac{\pi Q^4(\omega\pm2\Omega)^4}{3m^2c^4}\frac{|E_\pm/E|^2}{|d_\pm|^2}.
\nonumber 
\end{eqnarray}
The consistency of this model is corroborated by the fact that the sum of all partial cross sections ($\sigma_\omega+\sigma_{\omega+2\Omega}+\sigma_{\omega-2\Omega}+\sigma_{\rm mech}+\sigma_{\rm abs}$) equals the total extinction cross section derived from the optical theorem (see Appendix).

The partial cross sections are also resonant under the condition (\ref{cond}) (see Fig.\ \ref{Fig2}). In particular, light absorption [Fig.\ \ref{Fig2}(b)] produces a positive transfer of intrinsic angular momentum from each absorbed photon to the particle ($\propto\sigma_{\rm abs}I$), while elastic scattering [Fig.\ \ref{Fig2}(b)] dominates the $\Omega\sim\omega$ region. For right CP light, we obtain similar results, with the ellipsoids of $M$, $\sigma_{\rm abs}$, and $\sigma_\omega$ reflected with respect to the $\omega=0$ axis (not shown). For linear polarization, we find a superposition of the two orthogonal circular polarizations (see Appendix).

Two remarkable effects emerge at large rotation velocities ($|\Omega|>|\omega|$): (i) the inelastic emission (at frequency $\omega-\Omega$ for LCP light) becomes a leading process; and (ii) like the mechanical cross section, the total extinction cross section can be negative (e.g., for $\Omega>\omega>0$ and LCP light), thus confirming a net stimulated light emission, whereby  mechanical motion is converted into photons (see below).

\section{Connection to actual particles}

The model parameters $Q^2/m$, $\omega_0$, and $\gamma$ can be easily adjusted to fit the polarizability of a particle at rest ($\Omega=0$). Computing the induced dipole $\pb=Q\rho\rhoh$ from our model (see appendix), we find $p=\alpha E$, where $\alpha=(Q^2/m)/(\omega_0^2-\omega(\omega+\ii\gamma)-\ii\tau\omega^3)$ is the polarizability. In the absence of internal friction ($\gamma=0$), this expression satisfies the property ${\rm Im}\{-\alpha^{-1}\}=2\omega^3/3c^3$, as expected from the optical theorem for non-absorbing particles \cite{V1981}. This polarizability has the same form as that of a sphere of radius $R$ described by the Drude dielectric function $\epsilon=1-\omega_p^2/\omega(\omega+i\gamma)$, which allows us to identify $\omega_0=\omega_p/\sqrt{3}$ and $Q^2/m=\omega_0^2R^3$. Incidentally, at low $\Omega$ and within the dipole approximation, we find $M/|E|^2={\rm Im}\{\alpha_\parallel+\alpha_\perp\}-(4\omega^3/3c^3){\rm Re}\{\alpha_\parallel\alpha_\perp^*\}$ from an analysis based upon the Maxwell stress tensor \cite{J99,paper089} for the torque produced by LCP light on a rod under the conditions of Fig.\ \ref{Fig1}, where $\alpha_\parallel$ and $\alpha_\perp$ are the polarizabilities along directions parallel and perpendicular to the rod, respectively. This torque vanishes for non-absorbing spheres in virtue of the optical theorem. In contrast, non-absorbing rods ($\alpha_\parallel\neq\alpha_\perp$) experience a net torque because the term in $\omega^3/c^3$ cannot compensate the first one.

This indicates that the torque acting on small lossy spheres, in which absorption (the $\gamma$ term) governs ${\rm Im}\{-\alpha^{-1}\}$, can be approximated with our formalism as twice the torque acting on a rod, but using the above identification of model parameters. Just to give a better idea of the order of magnitude of the mechanical and absorption cross sections for a Drude sphere, the units in the color scale of Fig.\ \ref{Fig2}(a)-(b) are $Q^2\gamma/mc\omega_0^2=R^3/(c/\gamma)$, with typical values of $c/\gamma\sim1\,\mu$m in noble metals.

\section{Stimulated emission and light amplification}

\begin{figure}
\begin{center}
\includegraphics[width=85mm,angle=0,clip]{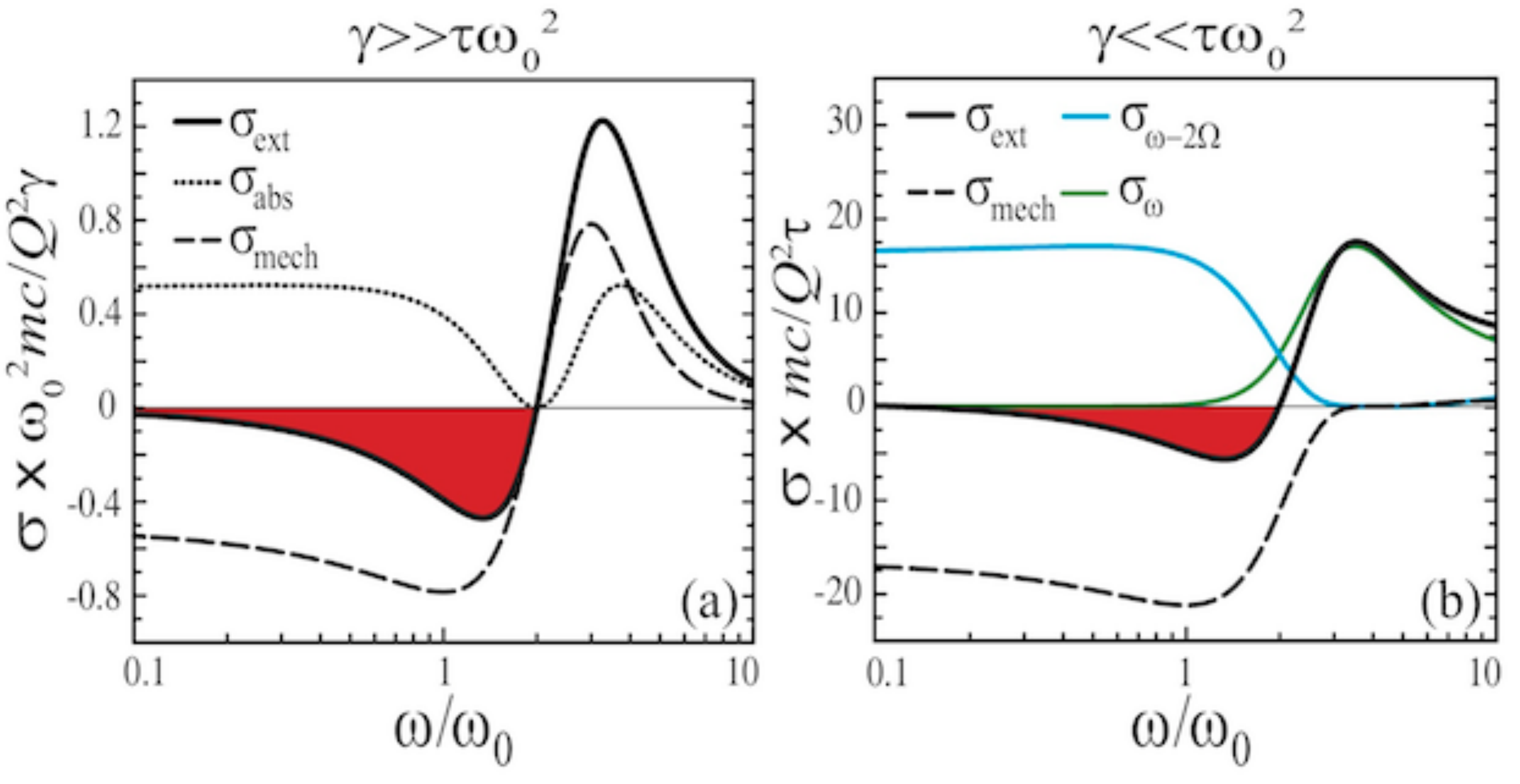}
\caption{Stimulated light emission in rotating particles. The plots show the spectral dependence of the partial cross sections for {\bf (a)} dissipative ($\gamma\gg\tau\omega_0^2$) and {\bf (b)} non-dissipative ($\gamma\ll\tau\omega_0^2$) rods rotating with frequency $\Omega=2\omega_0$. The shaded areas for negative extinction ($\sigma_{\rm ext}<0$) give the excess of photons added to the incident beam as a result of the interaction with the particle. We take $\gamma=0.1\,\omega_0$.} \label{Fig3}
\end{center}
\end{figure}

We show in Fig.\ \ref{Fig3} the partial cross sections of rotating rods in the limits of high and low ohmic losses. The negative extinction represents an increase in the amplitude of the incident beam after interaction with the rotating particle. Incidentally, our recently reported rotational friction \cite{pap157} can be understood as the spontaneous-emission counterpart of the stimulated emission under discussion. Stimulated emission takes place at $\omega<\Omega$ by transferring mechanical energy from the particle to the incident beam (coherently added photons). This allows us to speculate with the possibility of constructing a laser (see Appendix) in which an incident CP light beam is exponentially building up as it encounters rotating particles along its path (the particles are externally driven, for example, by pumping CP light of higher frequency). This scheme is robust because it is insensitive to finite distributions of particle size and rotation velocity, as long as the latter exceeds the light frequency.

\section{Particle heating}

\begin{figure}
\begin{center}
\includegraphics[width=70mm,angle=0,clip]{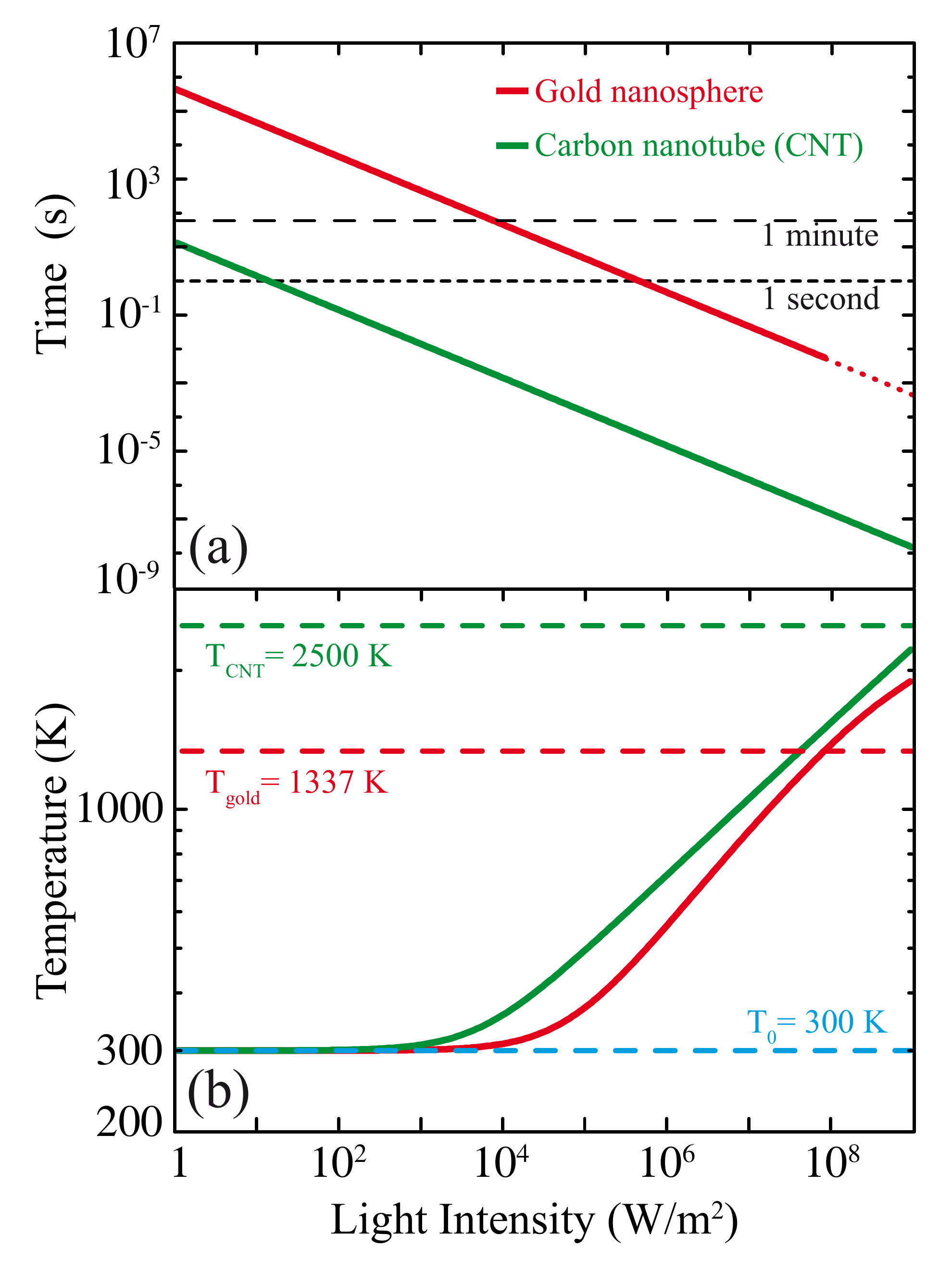}
\caption{{\bf (a)} Time needed to accelerate a 20-nm gold sphere and a $6\times6$ 10-nm-long single-wall carbon nanotube (CNT) to a rotational velocity $\Omega=1\,$MHz as a function of incident CP light intensity. {\bf (b)} Particle equilibrium temperature. The particles are rotating in vacuum at an external temperature $T_0=300\,$K. The gold and carbon melting points are indicated by dashed lines in (b). The light frequency is $\omega=2\pi\times10^{14}\,$Hz (wavelength $\sim3\,\mu$m).} \label{Fig4}
\end{center}
\end{figure}

For constant incident light intensity, light absorption leads to heating of the particle until it reaches an equilibrium temperature $T_{\rm eq}$ above the temperature of the surrounding vacuum $T_0$. This equilibrium is established when the absorbed power $\sigma_{\rm abs}I$ equals the radiative cooling rate $P^{\rm rad}$. In the Drude approximation for a metal sphere, and neglecting the effect of $\Omega$, which only enters at large velocities, we have \cite{pap157} $P^{\rm rad}\propto T_{\rm eq}^6-T_0^6$, 
which leads to $T_{\rm eq}^6=T_0^6+C\sigma_{\rm abs}I$. This is represented in Fig.\ \ref{Fig4}(b) for a 10\,nm carbon nanotube (upper solid curve) for which the polarizability has been obtained in the discrete-dipole approximation \cite{pap149}. There are two different regimes in the dependence of $T_{\rm eq}$ on light intensity: at low $I$, the particle is nearly at the vacuum temperature, but when the intensity increases, the particle heats up, asymptotically approaching a $I^{1/6}$ power law. A similar behavior is observed for a gold nanoparticle [Fig.\ \ref{Fig4}(b), lower solid curve], incorporating the effect of magnetic polarization and a realistic metal permittivity taken from optical data \cite{JC1972,pap157}.

\section{Rotational dynamics}

The time needed to accelerate a carbon nanotube and a gold nanoparticle to 1\,MHz is represented in Fig.\ \ref{Fig4}(a) as a function of LCP light intensity, as calculated with the torque of Eq.\ (\ref{M}). This time scales approximately as $\propto R^2\omega_0^2\Omega/\gamma\omega I$ under the condition $\gamma\omega\ll\omega_0^2$, and it plunges well under 0.01\,s for light intensities below the melting threshold. This suggests the possibility of achieving extreme rotation velocities in optically trapped nanoparticles, and eventually producing centrifugal explosion, thus introducing an unprecedented physical scenario (this should happen close to the point at which the centrifugal energy reaches the surface-tension energy, which in a 20-nm liquid gold particle occurs at $\Omega\sim3$GHz).




\section{Concluding remarks}

The present self-consistent oscillator model permits capturing the optical response of particles rotating under extremely high velocities. This defines a new scenario plagued with exotic phenomena such as strong inelastic scattering, and most notably, the possibility of realizing a laser (see Appendix) running on a classical linear system by exploiting our prediction of light amplification stimulated by particles rotating faster than the light frequency. There are several avenues towards potentially practical implementations of these systems, such as, for example, (1) a diluted gas of linear molecules driven to THz rotational frequencies by CP light and acting as the active medium of a laser operating in that demanded frequency range; (2) an extension of these ideas to acoustic lasing based on a similar classical interaction with subwavelength particles rotating faster than the sound frequency; (3) microwave waveguide setups such as those used to monitor rotational frequency shifts \cite{A1966}; (4) dust clouds in cosmic environments exposed to polarized light, in which radio wave amplification might be taking place.

\acknowledgments

This work has been supported by the Spanish MICINN (MAT2010-14885 and Consolider NanoLight.es) and the European Commission (FP7-ICT-2009-4-248855-N4E). A.A.-G. and A.M. acknowledge financial support through FPU from the Spanish ME.

\appendix

\section{Theoretical formalism}

\subsection{Description of the model and equation of motion}

\begin{figure}
\begin{center}
\includegraphics[width=80mm,angle=0,clip]{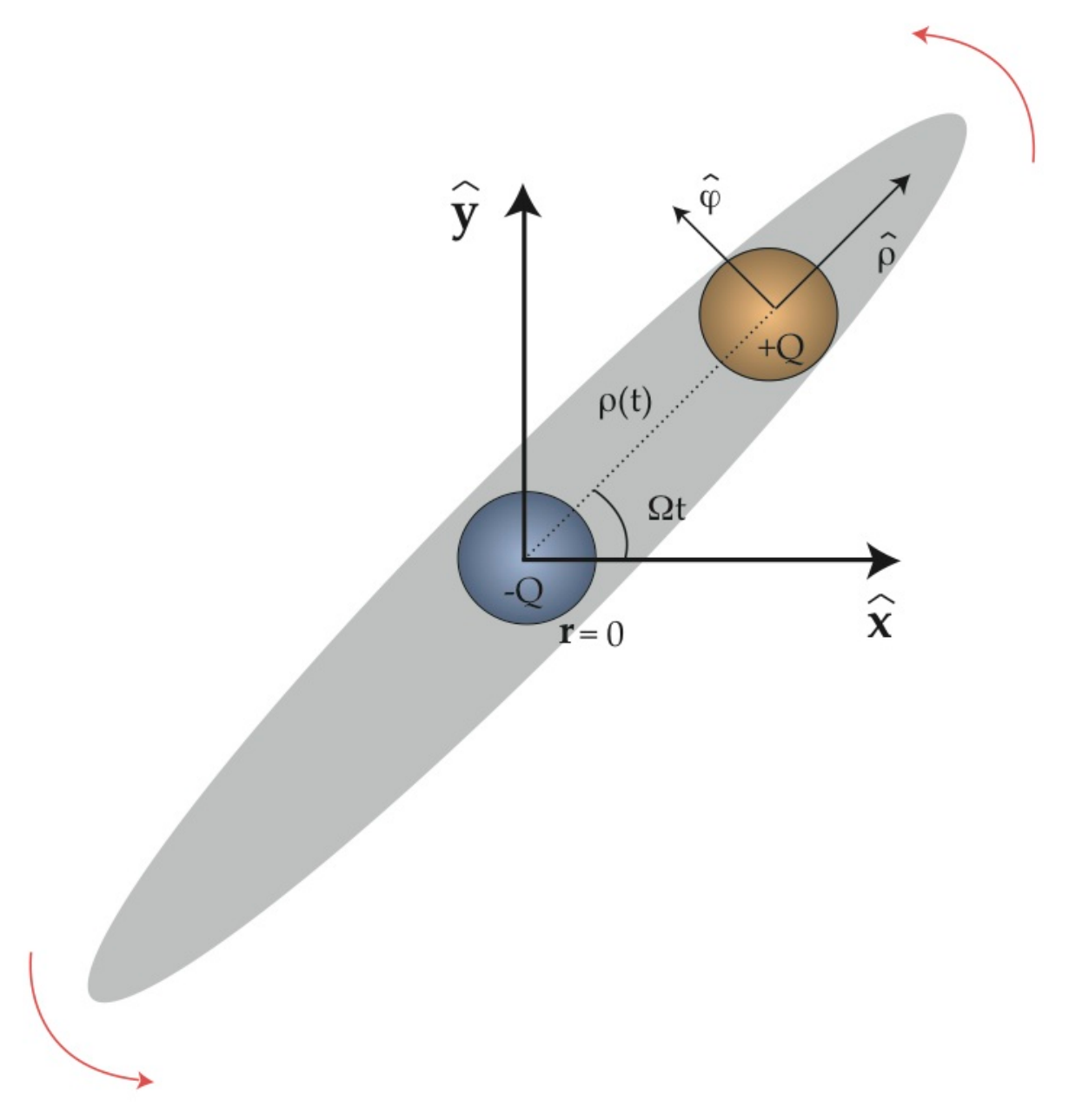}
\caption{Illustration of our model for a rotating nanorod. A charge $Q$ rotates with frequency $\Omega$ around a static charge $-Q$. The moving charge oscillates along the rod axis.} \label{FigSI1}
\end{center}
\end{figure}

We model a rotating rod by a point charge $Q$ oscillating around a fixed charge $-Q$ according to the equation of motion
\begin{eqnarray}
m\ddot{\rb}=-m\omega_0^2\rho\rhoh-m\gamma\dot{\rho}\rhoh+Q\Eb+m\tau\dddot{\rb}+F^{\rm react}\phih,
\label{force}
\end{eqnarray}
where $\Eb$ is the external electric field, the dots indicate differentiation with respect to the time, $\rb$ and $m$ are the coordinate vector and the mass of the moving charge, respectively, $\omega_0$ and $\gamma$ are the natural frequency and the intrinsic damping rate of the oscillator,
\[\tau=2Q^2/3mc^3\]
gives the coupling to radiation via an Abraham-Lorentz force \cite{J99}, and
\begin{eqnarray}
F^{\rm react}=\left(m\ddot{\rb}-Q\Eb-m\tau\dddot{\rb}\right)\cdot\phih
\label{Freact}
\end{eqnarray}
is the reaction of the nanoparticle that constraints the point charge to rotate with fixed frequency $\Omega$ around the $\zz$ axis. The point charge moves along the rod axis $\rhoh$, which also rotates with frequecy $\Omega$ around $\zz$. The main elements of the model are graphically illustrated in Fig.\ \ref{FigSI1}. The moving charge coordinate vector can thus be written as
\begin{eqnarray}
\rb(t)=\rho(t)\rhoh.
\label{r}
\end{eqnarray}
The radial and azimuthal unit vectors depend on time as $\rhoh=\xx\cos\Omega t+\yy\sin\Omega t$ and $\phih=-\xx\sin\Omega t+\yy\cos\Omega t$,
respectively. Inserting Eq.\ (\ref{r}) into Eq.\ (\ref{force}) and noticing that $\dot{\rhoh}=\Omega\phih$ and $\dot{\phih}=-\Omega\rhoh$, the radial equation of motion reduces to
\begin{eqnarray}
m\left(\omega_0^2\rho-\Omega^2\rho+\ddot{\rho}+\gamma\dot{\rho}
-\tau\dddot{\rho}+3\tau\Omega^2\dot{\rho}\right)=Q\Eb\cdot\rhoh,
\label{forcer}
\end{eqnarray}
whereas the azimuthal equation is trivially satisfied by our choice of $F^{\rm react}$.

We now write an external monochromatic electric field as $\Eb=(E_x\xx+E_y\yy)\ee^{-\ii\omega t}+c.c.$, which permits us to obtain for the inhomogeneous solution of Eq.\ (\ref{forcer}) the expression
\begin{eqnarray}
\rho(t)=\frac{Q}{2m}\left(\frac{E_+}{d_+}\ee^{-\ii\omega_+t}+
\frac{E_-}{d_-}\ee^{-\ii\omega_-t}+c.c.\right),
\label{rho}
\end{eqnarray}
where
\[\omega_\pm=\omega\pm\Omega,\]
\[E_\pm=E_x\pm\ii E_y,\]
and \[d_\pm=\omega_0^2-\Omega^2-\omega_\pm(\omega_\pm+\ii\gamma)-\ii\tau\omega_\pm(\omega_\pm^2+3\Omega^2).\]
Notice that the natural oscillation frequency $\omega_0$ is split into two shifted frequencies by the effect of a finite $\Omega$.

\subsection{Torque, absorption, and scattering cross section}

The torque acting on the particle is directed along $\zz$ and produced by the reaction force of Eq.\ (\ref{Freact}) according to $M=-F^{\rm react}\rho$. Inserting Eqs.\ (\ref{Freact}) and (\ref{rho}) into this expression and retaining only time-independent terms (i.e., the time-averaged contribution), we find
\begin{widetext}
\begin{eqnarray}
M=\frac{Q^2}{2m}\left[-\left(\gamma\omega_++\tau\omega_{++}^3\right)
\frac{|E_+|^2}{|d_+|^2}+\left(\gamma\omega_-+\tau\omega_{--}^3\right)
\frac{|E_-|^2}{|d_-|^2}\right],
\label{M}
\end{eqnarray}
\end{widetext}
where $\omega_{++}=\omega+2\Omega$ and $\omega_{--}=\omega-2\Omega$. Equation\ (\ref{M}) here is the same as Eq.\ (2) of the main paper.

The light exerts a mechanical torque on the particle resulting in a power transfer given by $P^{\rm mech}=M\Omega$. The particle is also capable of dissipating energy via light absorption by the material of which it is made. The total (mechanical plus absorption) work exerted by the external light on the particle per unit time has to include the effect of radiation damping, and it is therefore given by $P=\left(Q\Eb+m\tau\dddot{\rb}\right)\cdot\dot{\rb}$. Alternatively, we can write it as the power due to the dissipative force, $P=\left(m\gamma\dot{\rho}\rhoh\right)\cdot\dot{\rb}$. Using either one of these expressions, the time-averaged absorption power is found to reduce to
\begin{eqnarray}
P^{\rm abs}&=&P-P^{\rm mech}\nonumber\\&=&\frac{\gamma Q^2}{2m}\left(\omega_+^2
\frac{|E_+|^2}{|d_+|^2}+\omega_-^2\frac{|E_-|^2}{|d_-|^2}\right).
\nonumber
\end{eqnarray}
It is reassuring to observe that our model predicts the absorption power to be proportional to the intrinsic friction rate $\gamma$.
This result can be divided by the external light intensity $I=(c/2\pi)|E|^2$ to obtain the absorption cross section
\begin{eqnarray}
\sigma_{\rm abs}=\frac{\pi\gamma Q^2}{mc}\left(\omega_+^2
\frac{|E_+/E|^2}{|d_+|^2}+\omega_-^2\frac{|E_-/E|^2}{|d_-|^2}\right),
\label{sigmaabs}
\end{eqnarray}
where $|E|^2=|E_x|^2+|E_y|^2$.

\begin{figure*}
\begin{center}
\includegraphics[width=180mm,angle=0,clip]{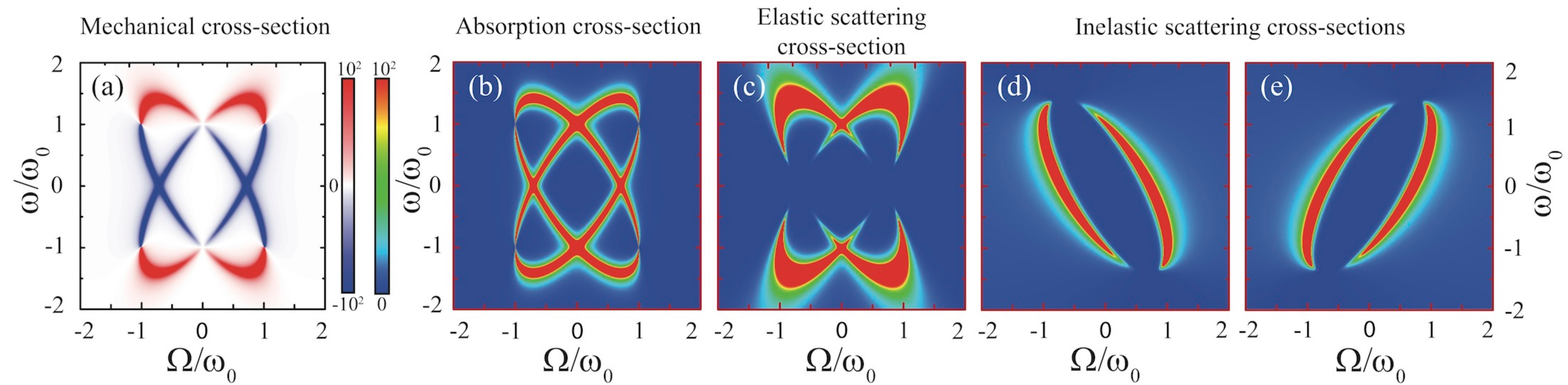}
\caption{Dynamical properties of a rotating rod for linearly polarized incident light of frequency $\omega$ and rotation frequency $\Omega$, both normalized to the excitation resonance frequency of the particle $\omega_0$. {\bf (a)} Mechanical cross section $M\Omega/I$, where $M$ is the electromagnetic torque and $I$ is the light intensity. {\bf (b)} Absorption cross-section. {\bf (c)} Elastic scattering cross-section. {\bf (d)} Inelastic scattering cross-section at frequency $\omega+2\Omega$. {\bf (e)} Inelastic scattering cross-section at frequency $\omega-2\Omega$. The particle is described by a rotating harmonic oscillator (Fig.\ \ref{FigSI1}) with intrinsic damping rate $\gamma=0.1\omega_0$. The color scale is in units of $Q^2\gamma/mc\omega_0^2$ in (a)-(b) and $Q^2\tau/mc$ in (c)-(e), assuming $\gamma\gg\tau\omega_0^2$, and it is saturated at the limits specified in the legend.} \label{FigSI2}
\end{center}
\end{figure*}

It is interesting to analyze the dipole induced in the rotating particle. We can write it as
\begin{widetext}
\begin{eqnarray}
{\bf p}=Q\rho\rhoh&=&\frac{Q^2}{4m}\Bigg\{\left[\left(\frac{E_+}{d_+}+\frac{E_-}{d_-}\right)\ee^{-\ii\omega t}+\frac{E_+}{d_+}\ee^{-\ii\omega_{++} t}+\frac{E_-}{d_-}\ee^{-\ii\omega_{--} t}\right]\xx
\nonumber\\&&\;\;\;\;\;-\ii\left[\left(\frac{E_+}{d_+}-\frac{E_-}{d_-}\right)\ee^{-\ii\omega t}-\frac{E_+}{d_+}\ee^{-\ii\omega_{++} t}+\frac{E_-}{d_-}\ee^{-\ii\omega_{--} t}\right]\yy\Bigg\}+c.c.
\label{dipole}
\end{eqnarray}
\end{widetext}
in the rest frame. This expression contains terms of frequencies $\omega$, $\omega+2\Omega$, and $\omega-2\Omega$. The elastic cross section corresponds to the field re-radiated by the component of frequency $\omega$,
\begin{eqnarray}
\sigma_\omega=\frac{\pi Q^4\omega^4}{3m^2c^4}\left(\frac{|E_+/E|^2}{|d_+|^2}+\frac{|E_-/E|^2}{|d_-|^2}\right),
\label{sigmaw}
\end{eqnarray}
as obtained from the time-averaged integral of the far-field Poynting vector (that is, the integral of the squared far-field amplitude) divided by the external field intensity $I$. Likewise, we can define inelastic cross sections corresponding to the emission of light with frequencies $\omega\pm2\Omega$,
\begin{eqnarray}
\sigma_{\omega\pm2\Omega}=\frac{\pi Q^4(\omega\pm2\Omega)^4}{3m^2c^4}\frac{|E_\pm/E|^2}{|d_\pm|^2}.
\label{sigmawpm}
\end{eqnarray}
The quantities $\sigma_{\omega\pm2\Omega}$ describe the effective area of the particles to inelastically scatter the incident light. These cross sections are the same as Eqs.\ (3)-(5) of the main paper. Numerical results are given there for left-handed circularly polarized incident light ($|E_+/E|^2=0$ and $|E_-/E|^2=2$) and also here in Fig.\ \ref{FigSI2} for linearly polarized light ($|E_+/E|^2=|E_-/E|^2=1$).

For the particle at rest ($\Omega=0$), with the electric field and the rod both oriented along $\xx$, the induced dipole of Eq.\ (\ref{dipole}) reduces to $(Q^2E/md)\ee^{-\ii\omega t}+c.c.$, where $d=\omega_0^2-\omega(\omega+\ii\gamma)-\ii\tau\omega^3$. The polarizability of the static particle is then given by $\alpha=Q^2/md$. Interestingly, this expression yields ${\rm Im}\{-1/\alpha\}=2\omega^3/3c^3+m\gamma\omega/Q^2$, in agreement with the optical theorem \cite{V1981}, which predicts ${\rm Im}\{-1/\alpha\}=2\omega^3/3c^3$ for a non-absorbing particle ($\gamma=0$).

The static polarizability permits establishing a connection between our simple oscillator model and actual physical parameters of real nanoparticles. For example, for a metallic ellipsoid of volume $V$ described by a Drude dielectric function of bulk plasmon frequency $\omega_p$ and damping rate $\gamma$, the electrostatic polarizability, corrected for radiative losses \cite{FW1984}, reduces to the above expression with $Q^2/m=\omega_p^2V/4\pi$ and $\omega_0=\omega_p\sqrt{L}$, where $L$ is the depolarization factor that depends on the rod aspect ratio \cite{paper112}.

\subsection{Balance of energy}

The optical theorem allows us to use the $\omega$ component of the induced dipole ${\bf p}$ [Eq.\ (\ref{dipole})] to obtain the total extinction cross section
\begin{widetext}
\begin{eqnarray}
\sigma_{\rm ext}&=&\frac{2\pi\omega}{c|E|^2}\left\langle\pb\cdot\Eb\right\rangle \label{sigmatot}\\
&=&\frac{\pi Q^2\omega}{mc}\left(\omega_+\left[\gamma+\tau(\omega_+^2+3\Omega^2)\right]\frac{|E_+/E|^2}{|d_+|^2}+\omega_-\left[\gamma+\tau(\omega_-^2+3\Omega^2)\right]\frac{|E_-/E|^2}{|d_-|^2}\right),
\nonumber
\end{eqnarray}
where $\langle\rangle$ denotes the time average. Furthermore, the mechanical work on the rotational motion can be expressed as the cross section
\begin{eqnarray}
\sigma_{\rm mech}&=&M\Omega/I  \label{sigmamech}\\
&=&\frac{\pi Q^2\Omega}{mc}\left[-\left(\gamma\omega_++\tau\omega_{++}^3\right)
\frac{|E_+/E|^2}{|d_+|^2}+\left(\gamma\omega_-+\tau\omega_{--}^3\right)
\frac{|E_-/E|^2}{|d_-|^2}\right],
\nonumber
\end{eqnarray}
\end{widetext}
which is directly obtained from Eq.\ (\ref{M}). Then, the total balance of power must be zero, and this leads to the condition
\begin{eqnarray}
\sigma_{\rm ext}=\sigma_\omega+\sigma_{\omega+2\Omega}+\sigma_{\omega-2\Omega}+\sigma_{\rm mech}+\sigma_{\rm abs},
\nonumber
\end{eqnarray}
which is indeed satisfied by Eqs.\ (\ref{sigmaabs}) and (\ref{sigmaw})-(\ref{sigmamech}). Incidentally, $\sigma_{\rm ext}$ can be negative, for example when $\Omega>\omega>0$ and $E_+=0$ (left circularly polarized light), pointing to a net stimulated emission of light from the rotating particle.

\section{Results for linearly polarized incident light}

In Fig.\ \ref{FigSI2}, we provide the equivalent of Fig.\ 2 of the main paper, but for linearly polarized incident light instead of circularly polarized light. The right and left circularly polarized components of the incident light produce inelastic emission at frequencies $\omega+2\Omega$ and $\omega-2\Omega$, respectively, as shown in Fig.\ \ref{FigSI2}(d)-(e). The torque and absorbed power are the superposition of the contributions originating from these two components. Interestingly, these terms add up in the absorption but they cancel each other in the torque at $\Omega=0$.

\section{Lasing and electromagnetic amplication in systems of rotating particles}

It is clear from Eq.\ (\ref{sigmatot}) that the extinction cross section can take negative values when the light and the particle rotate in the same direction (e.g., for $E_+=0$ and $\Omega>0$),  provided the particle rotates faster than the light frequency ($\Omega>\omega$). This is illustrated in Fig.\ 3 of the main paper, which shows that a negative cross section is possible due to either absorption or radiative processes, depending on the magnitude of $\gamma$ relative to $\tau\omega_0^2$. A negative extinction indicates that more light is leaving the system than it impinges on it. That is, light amplification takes place due to the interaction with the rotating particle, and the extra light is coherent with the incident one. These are necessary conditions for making a laser.

An electromagnetic amplifier could be constructed based on a medium filled with rotating particles. A light beam with the right characteristics would be amplified at every encounter with one of the particles. As a result, the beam could be exponentially amplified as it propagates along this medium. Particle rotation can be sustained by external means (e.g., either mechanically for radio waves or via optical pumping at higher frequencies), so that the system acts as a transducer that converts the externally supplied energy into coherent radiation.

In a more sophisticated arrangement, we envision an electromagnetic cavity containing such an amplifying medium, in which the radiation is making several passes before leaving the cavity (e.g., in a parallel mirror configuration). The seed in this system can be provided either by external illumination or by intrinsic spontaneous emission within the system, which has been recently studied in detail \cite{paper166}.

This lasing and light amplification effects can be potentially achievable within various windows of the electromagnetic spectrum. For example, one can rely on mechanically or magnetically induced motion up to radio frequencies $<1\,$ MHz. Light-driven rotation of nanoparticles levitated in vacuum could be used for higher frequencies in the GHz regime before the particles break apart. Molecular rotators could also be employed up to THz frequencies, in a process resembling the rotational Raman effect. These regimes can be accompanied by intrinsic resonance frequencies of the rotating bodies relying on LC circuits at low frequencies or physically equivalent excitations such as Mie resonances, plasmons, and optical phonons.

Finally, dust clouds in cosmic environments provide a plausible scenario for electromagnetic amplification because they are formed by microparticles that are exposed to the effect of neighboring light sources (e.g., stars and galaxies), which are circularly polarized in some cases. This illumination acts over enormous periods of time (millions of years), thus potentially inducing large rotation in the particles in the range of GHz before they are destroyed by centrifugal explosion. Therefore, amplification can be expected up to GHz frequencies, compatible with the low-end of the cosmic microwave background.


\begin{thebibliography}{26}
\expandafter\ifx\csname natexlab\endcsname\relax\def\natexlab#1{#1}\fi
\expandafter\ifx\csname bibnamefont\endcsname\relax
  \def\bibnamefont#1{#1}\fi
\expandafter\ifx\csname bibfnamefont\endcsname\relax
  \def\bibfnamefont#1{#1}\fi
\expandafter\ifx\csname citenamefont\endcsname\relax
  \def\citenamefont#1{#1}\fi
\expandafter\ifx\csname url\endcsname\relax
  \def\url#1{\texttt{#1}}\fi
\expandafter\ifx\csname urlprefix\endcsname\relax\def\urlprefix{URL }\fi
\providecommand{\bibinfo}[2]{#2}
\providecommand{\eprint}[2][]{\url{#2}}

\bibitem[{\citenamefont{Ashkin}(1970)}]{A1970}
\bibinfo{author}{\bibfnamefont{A.}~\bibnamefont{Ashkin}},
  \bibinfo{journal}{Phys.\ Rev.\ Lett.} \textbf{\bibinfo{volume}{24}},
  \bibinfo{pages}{156} (\bibinfo{year}{1970}).

\bibitem[{\citenamefont{Grier}(2003)}]{G03}
\bibinfo{author}{\bibfnamefont{D.~G.} \bibnamefont{Grier}},
  \bibinfo{journal}{Nature} \textbf{\bibinfo{volume}{424}},
  \bibinfo{pages}{810} (\bibinfo{year}{2003}).

\bibitem[{\citenamefont{Groeblacher et~al.}(2009)\citenamefont{Groeblacher,
  Hertzberg, Vanner, Cole, Gigan, Schwab, and Aspelmeyer}}]{GHV09}
\bibinfo{author}{\bibfnamefont{S.}~\bibnamefont{Groeblacher}},
  \bibinfo{author}{\bibfnamefont{J.~B.} \bibnamefont{Hertzberg}},
  \bibinfo{author}{\bibfnamefont{M.~R.} \bibnamefont{Vanner}},
  \bibinfo{author}{\bibfnamefont{G.~D.} \bibnamefont{Cole}},
  \bibinfo{author}{\bibfnamefont{S.}~\bibnamefont{Gigan}},
  \bibinfo{author}{\bibfnamefont{K.~C.} \bibnamefont{Schwab}},
  \bibnamefont{and}
  \bibinfo{author}{\bibfnamefont{M.}~\bibnamefont{Aspelmeyer}},
  \bibinfo{journal}{Nat.\ Phys.} \textbf{\bibinfo{volume}{5}},
  \bibinfo{pages}{485} (\bibinfo{year}{2009}).

\bibitem[{\citenamefont{Beth}(1936)}]{B1936}
\bibinfo{author}{\bibfnamefont{R.~A.} \bibnamefont{Beth}},
  \bibinfo{journal}{Phys.\ Rev.} \textbf{\bibinfo{volume}{50}},
  \bibinfo{pages}{115} (\bibinfo{year}{1936}).

\bibitem[{\citenamefont{Allen}(1966)}]{A1966}
\bibinfo{author}{\bibfnamefont{P.~J.} \bibnamefont{Allen}},
  \bibinfo{journal}{AJP} \textbf{\bibinfo{volume}{23}}, \bibinfo{pages}{1185}
  (\bibinfo{year}{1966}).

\bibitem[{\citenamefont{Friese et~al.}(1998)\citenamefont{Friese, Nieminen,
  Heckenberg, and Rubinsztein-Dunlop}}]{FNH98}
\bibinfo{author}{\bibfnamefont{M.~E.~J.} \bibnamefont{Friese}},
  \bibinfo{author}{\bibfnamefont{T.~A.} \bibnamefont{Nieminen}},
  \bibinfo{author}{\bibfnamefont{N.~R.} \bibnamefont{Heckenberg}},
  \bibnamefont{and}
  \bibinfo{author}{\bibfnamefont{H.}~\bibnamefont{Rubinsztein-Dunlop}},
  \bibinfo{journal}{Nature} \textbf{\bibinfo{volume}{394}},
  \bibinfo{pages}{348} (\bibinfo{year}{1998}).

\bibitem[{\citenamefont{Liu et~al.}(2010)\citenamefont{Liu, Zentgraf, Liu,
  Bartal, and Zhang}}]{LZL10}
\bibinfo{author}{\bibfnamefont{M.}~\bibnamefont{Liu}},
  \bibinfo{author}{\bibfnamefont{T.}~\bibnamefont{Zentgraf}},
  \bibinfo{author}{\bibfnamefont{Y.}~\bibnamefont{Liu}},
  \bibinfo{author}{\bibfnamefont{G.}~\bibnamefont{Bartal}}, \bibnamefont{and}
  \bibinfo{author}{\bibfnamefont{X.}~\bibnamefont{Zhang}},
  \bibinfo{journal}{Nat.\ Nanotech.} \textbf{\bibinfo{volume}{5}},
  \bibinfo{pages}{570} (\bibinfo{year}{2010}).

\bibitem[{\citenamefont{Kane}(2010)}]{K10}
\bibinfo{author}{\bibfnamefont{B.~E.} \bibnamefont{Kane}},
  \bibinfo{journal}{Phys.\ Rev.\ B} \textbf{\bibinfo{volume}{82}},
  \bibinfo{pages}{115441} (\bibinfo{year}{2010}).

\bibitem[{\citenamefont{Gahagan and Swartzlander}(1998)}]{GS98}
\bibinfo{author}{\bibfnamefont{K.~T.} \bibnamefont{Gahagan}} \bibnamefont{and}
  \bibinfo{author}{\bibfnamefont{G.~A.} \bibnamefont{Swartzlander}},
  \bibinfo{journal}{J.\ Opt.\ Soc.\ Am.\ B} \textbf{\bibinfo{volume}{15}},
  \bibinfo{pages}{524} (\bibinfo{year}{1998}).

\bibitem[{\citenamefont{Galajda and Ormos}(2001)}]{GO01}
\bibinfo{author}{\bibfnamefont{P.}~\bibnamefont{Galajda}} \bibnamefont{and}
  \bibinfo{author}{\bibfnamefont{P.}~\bibnamefont{Ormos}},
  \bibinfo{journal}{Appl.\ Phys.\ Lett.} \textbf{\bibinfo{volume}{78}},
  \bibinfo{pages}{249} (\bibinfo{year}{2001}).

\bibitem[{\citenamefont{Tong et~al.}(2010)\citenamefont{Tong, Miljkovi\'c, and
  K\"all}}]{TMK10}
\bibinfo{author}{\bibfnamefont{L.}~\bibnamefont{Tong}},
  \bibinfo{author}{\bibfnamefont{V.~D.} \bibnamefont{Miljkovi\'c}},
  \bibnamefont{and} \bibinfo{author}{\bibfnamefont{M.}~\bibnamefont{K\"all}},
  \bibinfo{journal}{Nano\ Lett.} \textbf{\bibinfo{volume}{10}},
  \bibinfo{pages}{268} (\bibinfo{year}{2010}).

\bibitem[{\citenamefont{{I. Bialynicki-Birula} and {Z.
  Bialynicka-Birula}}(1997)}]{BB97}
\bibinfo{author}{\bibnamefont{{I. Bialynicki-Birula}}} \bibnamefont{and}
  \bibinfo{author}{\bibnamefont{{Z. Bialynicka-Birula}}},
  \bibinfo{journal}{Phys.\ Rev.\ Lett.} \textbf{\bibinfo{volume}{78}},
  \bibinfo{pages}{2539} (\bibinfo{year}{1997}).

\bibitem[{\citenamefont{Courtial
  et~al.}(1998{\natexlab{a}})\citenamefont{Courtial, Dholakia, Robertson,
  Allen, and Padgett}}]{CDR98}
\bibinfo{author}{\bibfnamefont{J.}~\bibnamefont{Courtial}},
  \bibinfo{author}{\bibfnamefont{K.}~\bibnamefont{Dholakia}},
  \bibinfo{author}{\bibfnamefont{D.~A.} \bibnamefont{Robertson}},
  \bibinfo{author}{\bibfnamefont{L.}~\bibnamefont{Allen}}, \bibnamefont{and}
  \bibinfo{author}{\bibfnamefont{M.~J.} \bibnamefont{Padgett}},
  \bibinfo{journal}{Phys.\ Rev.\ Lett.} \textbf{\bibinfo{volume}{80}},
  \bibinfo{pages}{3217} (\bibinfo{year}{1998}{\natexlab{a}}).

\bibitem[{\citenamefont{Courtial
  et~al.}(1998{\natexlab{b}})\citenamefont{Courtial, Robertson, Dholakia,
  Allen, and Padgett}}]{CRD98}
\bibinfo{author}{\bibfnamefont{J.}~\bibnamefont{Courtial}},
  \bibinfo{author}{\bibfnamefont{D.~A.} \bibnamefont{Robertson}},
  \bibinfo{author}{\bibfnamefont{K.}~\bibnamefont{Dholakia}},
  \bibinfo{author}{\bibfnamefont{L.}~\bibnamefont{Allen}}, \bibnamefont{and}
  \bibinfo{author}{\bibfnamefont{M.~J.} \bibnamefont{Padgett}},
  \bibinfo{journal}{Phys.\ Rev.\ Lett.} \textbf{\bibinfo{volume}{81}},
  \bibinfo{pages}{4828} (\bibinfo{year}{1998}{\natexlab{b}}).

\bibitem[{\citenamefont{Michalski et~al.}(2005)\citenamefont{Michalski,
  H\"uttner, and Schimming}}]{MHS05}
\bibinfo{author}{\bibfnamefont{M.}~\bibnamefont{Michalski}},
  \bibinfo{author}{\bibfnamefont{W.}~\bibnamefont{H\"uttner}},
  \bibnamefont{and}
  \bibinfo{author}{\bibfnamefont{H.}~\bibnamefont{Schimming}},
  \bibinfo{journal}{Phys.\ Rev.\ Lett.} \textbf{\bibinfo{volume}{95}},
  \bibinfo{pages}{203005} (\bibinfo{year}{2005}).

\bibitem[{\citenamefont{Ashkin and Dziedzic}(1976)}]{AD1976}
\bibinfo{author}{\bibfnamefont{A.}~\bibnamefont{Ashkin}} \bibnamefont{and}
  \bibinfo{author}{\bibfnamefont{J.~M.} \bibnamefont{Dziedzic}},
  \bibinfo{journal}{Appl.\ Phys.\ Lett.} \textbf{\bibinfo{volume}{28}},
  \bibinfo{pages}{333} (\bibinfo{year}{1976}).

\bibitem[{\citenamefont{Garetz}(1981)}]{G1981}
\bibinfo{author}{\bibfnamefont{B.~A.} \bibnamefont{Garetz}},
  \bibinfo{journal}{J.\ Opt.\ Soc.\ Am.\ Lett.} \textbf{\bibinfo{volume}{71}},
  \bibinfo{pages}{609} (\bibinfo{year}{1981}).

\bibitem[{\citenamefont{Jackson}(1999)}]{J99}
\bibinfo{author}{\bibfnamefont{J.~D.} \bibnamefont{Jackson}},
  \emph{\bibinfo{title}{Classical Electrodynamics}}
  (\bibinfo{publisher}{Wiley}, \bibinfo{address}{New York},
  \bibinfo{year}{1999}).

\bibitem[{\citenamefont{Chute}(1967)}]{C1967}
\bibinfo{author}{\bibfnamefont{F.~S.} \bibnamefont{Chute}},
  \bibinfo{journal}{IEEE\ Trans.\ Antennas\ Propag.}
  \textbf{\bibinfo{volume}{15}}, \bibinfo{pages}{585} (\bibinfo{year}{1967}).

\bibitem[{\citenamefont{{van de Hulst}}(1981)}]{V1981}
\bibinfo{author}{\bibfnamefont{H.~C.} \bibnamefont{{van de Hulst}}},
  \emph{\bibinfo{title}{Light Scattering by Small Particles}}
  (\bibinfo{publisher}{Dover}, \bibinfo{address}{New York},
  \bibinfo{year}{1981}).

\bibitem[{\citenamefont{Dennis et~al.}(2007)\citenamefont{Dennis, Zheludev, and
  {Garc\'{\i}a de Abajo}}}]{paper089}
\bibinfo{author}{\bibfnamefont{M.~R.} \bibnamefont{Dennis}},
  \bibinfo{author}{\bibfnamefont{N.~I.} \bibnamefont{Zheludev}},
  \bibnamefont{and} \bibinfo{author}{\bibfnamefont{F.~J.}
  \bibnamefont{{Garc\'{\i}a de Abajo}}}, \bibinfo{journal}{Opt.\ Express}
  \textbf{\bibinfo{volume}{15}}, \bibinfo{pages}{9692} (\bibinfo{year}{2007}).

\bibitem[{\citenamefont{Manjavacas and {Garc\'{\i}a de
  Abajo}}(2010{\natexlab{a}})}]{pap157}
\bibinfo{author}{\bibfnamefont{A.}~\bibnamefont{Manjavacas}} \bibnamefont{and}
  \bibinfo{author}{\bibfnamefont{F.~J.} \bibnamefont{{Garc\'{\i}a de Abajo}}},
  \bibinfo{journal}{Phys.\ Rev.\ Lett.} \textbf{\bibinfo{volume}{105}},
  \bibinfo{pages}{113601} (\bibinfo{year}{2010}{\natexlab{a}}).

\bibitem[{\citenamefont{Johnson and Christy}(1972)}]{JC1972}
\bibinfo{author}{\bibfnamefont{P.~B.} \bibnamefont{Johnson}} \bibnamefont{and}
  \bibinfo{author}{\bibfnamefont{R.~W.} \bibnamefont{Christy}},
  \bibinfo{journal}{Phys.\ Rev.\ B} \textbf{\bibinfo{volume}{6}},
  \bibinfo{pages}{4370} (\bibinfo{year}{1972}).

\bibitem[{\citenamefont{Ford and Weber}(1984)}]{FW1984}
\bibinfo{author}{\bibfnamefont{G.~W.} \bibnamefont{Ford}} \bibnamefont{and}
  \bibinfo{author}{\bibfnamefont{W.~H.} \bibnamefont{Weber}},
  \bibinfo{journal}{Phys.\ Rep.} \textbf{\bibinfo{volume}{113}},
  \bibinfo{pages}{195} (\bibinfo{year}{1984}).

\bibitem[{\citenamefont{Myroshnychenko
  et~al.}(2008)\citenamefont{Myroshnychenko, {Rodr\'{\i}guez-Fern\'andez},
  Pastoriza-Santos, Funston, Novo, Mulvaney, {Liz-Marz\'an}, and {Garc\'{\i}a
  de Abajo}}}]{paper112}
\bibinfo{author}{\bibfnamefont{V.}~\bibnamefont{Myroshnychenko}},
  \bibinfo{author}{\bibfnamefont{J.}~\bibnamefont{{Rodr\'{\i}guez-Fern\'andez}%
}}, \bibinfo{author}{\bibfnamefont{I.}~\bibnamefont{Pastoriza-Santos}},
  \bibinfo{author}{\bibfnamefont{A.~M.} \bibnamefont{Funston}},
  \bibinfo{author}{\bibfnamefont{C.}~\bibnamefont{Novo}},
  \bibinfo{author}{\bibfnamefont{P.}~\bibnamefont{Mulvaney}},
  \bibinfo{author}{\bibfnamefont{L.~M.} \bibnamefont{{Liz-Marz\'an}}},
  \bibnamefont{and} \bibinfo{author}{\bibfnamefont{F.~J.}
  \bibnamefont{{Garc\'{\i}a de Abajo}}}, \bibinfo{journal}{Chem.\ Soc.\ Rev.}
  \textbf{\bibinfo{volume}{37}}, \bibinfo{pages}{1792} (\bibinfo{year}{2008}).

\bibitem[{\citenamefont{Manjavacas and {Garc\'{\i}a de
  Abajo}}(2010{\natexlab{b}})}]{paper166}
\bibinfo{author}{\bibfnamefont{A.}~\bibnamefont{Manjavacas}} \bibnamefont{and}
  \bibinfo{author}{\bibfnamefont{F.~J.} \bibnamefont{{Garc\'{\i}a de Abajo}}},
  \bibinfo{journal}{Phys.\ Rev.\ A} \textbf{\bibinfo{volume}{82}},
  \bibinfo{pages}{063827} (\bibinfo{year}{2010}{\natexlab{b}}).

\end{thebibliography}

\end{document}